\title{The Shannon capacity of a graph and the independence numbers of its powers}
\author{{Noga Alon\thanks{ Schools of Mathematics and Computer Science,
Raymond and Beverly Sackler Faculty of Exact Sciences, Tel Aviv
University, Tel Aviv, 69978, Israel. Email: nogaa@tau.ac.il.
Research supported in part by a USA-Israeli BSF grant, by the
Israel Science Foundation and by the Hermann Minkowski Minerva
Center for Geometry at Tel Aviv University.}} \quad {Eyal Lubetzky
\thanks{ School of Computer Science, Raymond and Beverly
Sackler Faculty of Exact Sciences, Tel Aviv University, Tel Aviv,
69978, Israel. Email: lubetzky@tau.ac.il. Research partially
supported by a Charles Clore Foundation Fellowship.}}}

\documentclass [11pt]{article}
\usepackage[]{epsf,epsfig,amsmath,amssymb,amsfonts,latexsym,amsthm}

\newtheorem{theorem}{Theorem}

\newtheorem{conjecture}[theorem]{Conjecture}

\newtheorem{question}[theorem]{Question}

\renewcommand{\epsilon}{\varepsilon}

\begin{document}
\maketitle

\begin{abstract}
The independence numbers of powers of graphs have been long studied,
under several definitions of graph products, and in particular,
under the strong graph product. We show that the series of
independence numbers in strong powers of a fixed graph can exhibit a
complex structure, implying that the Shannon Capacity of a graph
cannot be approximated (up to a sub-polynomial factor of the number
of vertices) by any arbitrarily large, yet fixed, prefix of the
series. This is true even if this prefix shows a significant
increase of the independence number at a given power, after which it
stabilizes for a while.
\end{abstract}

\section{Introduction}
\label{sec::intro} Given two graphs, $G_1$ and $G_2$, their
\textit{strong graph product} $G_1 \cdot G_2$ has a vertex set
$V(G_1) \times V(G_2)$, and two distinct vertices $(v_1,v_2)$ and
$(u_1,u_2)$ are connected iff they are adjacent or equal in each
coordinate (i.e., for $i \in \{1,2\}$, either $v_i=u_i$ or $v_i
u_i \in E(G_i)$). This product is associative and commutative, and
we can thus define $G^k$ as the product of $k$ copies of $G$. In
\cite{Shannon}, Shannon introduced the parameter $c(G)$, the
\textit{Shannon Capacity} of a graph $G$, which is the limit
$\lim_{k \rightarrow \infty}\sqrt[k]{\alpha(G^k)}$, where
$\alpha(G^k)$ is the independence number of $G^k$ (it is easy to
see that this limit exists by super-multiplicativity). The
considerable amount of interest that $c(G)$ has received (see,
e.g., \cite{NogaRamsey}, \cite{NogaUnion}, \cite{Bohman},
\cite{BohmanHolzman}, \cite{Haemers}, \cite{HaemersOnLovasz},
\cite{LovaszTheta}, \cite{Schrijver}, \cite{VanLintWilson}) is
motivated by Information Theory concerns: this parameter
represents the effective size of an alphabet, in a communication
model where the graph $G$ represents the channel. In other words,
we consider a transmission scheme where the input is a set of
single letters $V=\{1,\ldots,n\}$, and our graph $G$ has $V$ as
its set of vertices, and an edge between each pair of letters, iff
they are confusable in transmission (i.e., $(1,2)\in E(G)$
indicates that sending an input of $1$ or an input of $2$ might
result in the same output). Clearly $\alpha(G)$ is the maximum
size of a set of single letters which can be predefined, then sent
with zero-error. By definition, $\alpha(G^k)$ represents such a
set of words of length $k$ (since two distinct words are
distinguishable iff at least one of their coordinates is
distinguishable), leading to the intuitive interpretation of
$c(G)$ as the effective size of the alphabet of the channel
(extending the word length to infinity, while normalizing it in
each step).

\noindent Consider the series $a_k = a_k(G) =
\sqrt[k]{\alpha(G^k)}$, which we call "the \textit{independence
series} of $G$". As observed in \cite{Shannon}, the limit
$c(G)=\lim_{k \rightarrow \infty}a_k$ exists and equals its
supremum, and $a_{m k} \geq a_k$ for all integers $m,k$. Our
motivation for the study of the series $a_k$ is the computational
problem of approximating $c(G)$. So far, all graphs whose Shannon
capacity is known, attain the capacity either at $a_1$ (the
independence number, e.g., perfect graphs), $a_2$ (e.g., self
complementary vertex-transitive graphs) or do not attain it at any
$a_k$ (e.g., the cycle $C_5$ with the addition of an isolated
vertex). One might suspect that once the $a_k$ series remains
roughly a constant for several consecutive values of $k$, its value
becomes a good approximation to its limit, $c(G)$. This, however, is
false. Moreover, it remains false even when restricting ourselves to
cases where $a_k$ increases significantly before it stabilizes for a
few steps. We thus address the following questions:
\begin{enumerate} \item Is it true that for every arbitrarily
large integer $k$, there is a $\delta = \delta(k) > 0$ and a graph
$G$ on $n$ vertices such that the values $\{a_i\}_{i<k}$ are all
at least $n^\delta$- far from $c(G)$? \item Can the series $a_k$
increase significantly (in terms of $n=|V(G)|$) in an arbitrary
number of places?
\end{enumerate}
In this short paper we show that the answer to both questions above
is positive. The first question is settled by Theorem
\ref{thm-one-jump}, proved in section \ref{sec::thm-one-proof}.
\begin{theorem}\label{thm-one-jump} For every
fixed $\nu \in \mathbb{N}$ and $\epsilon > 0$ there exists a graph
$G$ on $N$ vertices such that for all $k<\nu$, $a_k \leq c_k
\log_2(N)$ (where $c_k=c_{k,\nu}$), and yet $a_\nu \geq
N^{\frac{1}{\nu}}$.
\end{theorem}
\noindent Indeed, for any fixed $k$, there exists a graph $G$ on
$N$ vertices, whose Shannon Capacity satisfies $c(G)
> N^\delta \max_{i<k}\{a_i\}$, where $\delta=\frac{1-o(1)}{k}$.

\noindent Theorem \ref{thm-multiple-jumps}, proved in section
\ref{sec::thm-two-proof}, settles the second question, and implies
the existence of a graph $G$ whose independence series $a_k$
contains an arbitrary number of "jumps" at arbitrarily chosen
locations; hence, noticing a significant increase in this series, or
noticing that it stabilizes for a while, does not ensure any
proximity to $c(G)$.

\begin{theorem}\label{thm-multiple-jumps}For every fixed
$\nu_1 < \ldots < \nu_s \in \mathbb{N}$ and $\epsilon>0$ there
exists a graph $G$ such that for all $k<\nu_i$, $a_k <
a_{\nu_i}^\epsilon $ ($i\in\{1,\ldots,s\}$).\end{theorem}

The above theorems imply that the naive approach of computing the
$a_k$ values for some $k$ does not provide even a PSPACE algorithm
for approximating $c(G)$. Additional remarks on the complexity of
the problem of estimating $c(G)$, as well as several open problems,
appear in the final section \ref{sec::final}.

\section{The capacity and the initial $a_k$-s}
\label{sec::thm-one-proof} In this section we prove Theorem
\ref{thm-one-jump}, using a probabilistic approach, which is based
on the method of \cite{NogaRepeatedComm}, but requires some
additional ideas.

\noindent Let $2 \leq \nu \in \mathbb{N}$; define $N=n\nu $ ($n$
will be a sufficiently large integer) , and let
$V(G)=\{0,\ldots,N-1\}$. Let ${\cal R}$ denote the equivalence
relation on the set of unordered pairs of
\textit{distinct} vertices, in which $(x,y)$ is identical to
$(y,x)$ and is equivalent to $(x+k n,y+k n)$ for all $0 \leq k
\leq \nu-1$, where addition is reduced modulo $N$. Let $\{{\cal
R}_1,\ldots,{\cal R}_M\}$ denote the different equivalence classes
of ${\cal R}$. For every $x \neq y$, let ${\cal R}(x,y)$ denote
the equivalence class of $(x,y)$ under $\cal{R}$; then either
$|{\cal R}(x,y)|$ is precisely $\nu$, or the following equality
holds for some $l < \nu$:
$$ (x,y)\equiv(y+l n,x+l n) \pmod{N} $$
This implies that $N \mid 2 l n$, hence $2l = \nu$. We deduce that
if $\nu$ is odd, $|{\cal R}_i|=\nu$ for all $1\leq i\leq M$, and
$M=\frac{1}{\nu}\binom{N}{2}$. If $\nu$ is even:
$$ |{\cal R}(x,y)| =
\begin{cases}
    \frac{1}{2}\nu & \text{If } y \equiv x+\frac{1}{2}\nu n, \\
    \nu & \text{Otherwise}.
  \end{cases}
$$
and $M=\frac{1}{\nu}\binom{N}{2}+\frac{1}{2\nu}N$, i.e., in case of
an even $\nu$ there are $N/2$ pairs which belong to $n$ smaller
classes, each of which is of size $\frac{1}{2}\nu$, while the
remaining edges belong to ordinary edge classes of size $\nu$.

\noindent The edges of $G$ are chosen randomly, by starting with
the complete graph and excluding a single edge from each
equivalence class, uniformly and independently, thus
$|E(G)|=\binom{N}{2}-M = \binom{N}{2}\left(\frac{\nu-1}{\nu}+o(1)\right)$. \\
\noindent A standard first moment consideration (c.f., e.g.,
\cite{ProbMethod}) shows that $a_1 = \alpha(G) < \lceil 2\log_\nu(N)
\rceil$ almost surely. To see this, set $s=\lceil
2\log_\nu(N)\rceil$, and take an arbitrary set $S \subset V(G)$ of
size $s$. If $S$ contains more than one member of some edge class
${\cal R}_i$, it cannot be independent. Otherwise, its edge
probabilities are independent, and all that is left is examining the
lengths of the corresponding edge classes. Assume $S$ contains $r$
pairs which belong to short edge classes:
$(x_1,y_1),\ldots,(x_r,y_r)$. If $\nu$ is odd, $r=0$, otherwise $y_i
= x_i + \frac{1}{2}\nu n$ for all $i$, and $x_i \neq x_j
\pmod{\frac{1}{2}\nu n}$ for all $i \neq j$ (distinct pairs in $S$
belong to distinct edge classes). It follows that $r \leq
\frac{s}{2}$, and we deduce that for each such set $S$:
$$\Pr[\mbox{$S$ is independent}] \leq
\left(\frac{1}{\nu}\right)^{\binom{s}{2}-r}
\left(\frac{2}{\nu}\right)^r
 \leq \left(\frac{1}{\nu}\right)^{\binom{s}{2}} 2^{s/2}
$$ Applying a union bound and using the
fact that $\frac{(2\nu)^{s/2}}{s!}$ tends to 0 as $N$, and hence
$s$, tend to infinity, we obtain:
$$\Pr[\alpha(G) \geq s] \leq \binom{N}{s}\nu^{-\binom{s}{2}}
2^{s/2} \leq \frac{2^{s/2}}{s!}\left(N \nu^{-\frac{s-1}{2}}\right)^s
= \frac{(2\nu)^{s/2}}{s!}\left(N \nu^{-\frac{s}{2}}\right)^s \leq
\frac{(2\nu)^{s/2}}{s!} = o(1) ~,$$ where the $o(1)$ term here, and
in what follows, tends to 0 as $N$ tends to infinity.

\noindent We next deal with $G^k$ for $2\leq k<\nu$. Fix a set $S
\subset V(G^k)$ of size $s=\lceil c_k \log_2^k(N) \rceil$, where
$c_k$ will be determined later. Define $S'$, a subset of $S$, in the
following manner: start with $S'=\phi$, order the vertices of $S$
arbitrarily, and then process them one by one according to that
order. When processing a vertex $v=(v_1,\ldots,v_k)\in S$, we add it
to $S'$, and remove from $S$ all of the following vertices which
contain $v_i + t n \pmod{N}$ in any of their coordinates, for any
$i\in[k]$ and $t \in \{0,\ldots,\nu-1\}$. In other words, once we
add $v$ to $S'$, we make sure that its coordinates modulo $n$ will
not appear anywhere else in $S'$. If $S$ is independent, it has at
most $\alpha(G^{k-1})$ vertices with a fixed coordinate, thus
$s'=|S'| \geq s / \left(k^2 \cdot \nu \cdot \alpha(G^{k-1})\right)$.
Notice that each two distinct vertices $u,v \in S'$ have distinct
vertices of $G$ in every coordinate, thus ${\cal R}(u_i,v_i)$ is
defined for all $i$; furthermore, for any \textit{other} pair of
distinct vertices $u',v' \in S'$, the sets $\{{\cal
R}(u_1,v_1),\ldots,{\cal R}(u_k,v_k)\}$ and
$\{{\cal R}(u'_1,v'_1),\ldots,{\cal R}(u'_k,v'_k)\}$ are disjoint. \\
\noindent We next bound the probability of an edge between a pair
of vertices $u \neq v \in S'$. Let $k'$ denote the number of
\textbf{distinct} pairs of corresponding coordinates of $u,v$, and
let $t_l$, $1\leq l \leq M$, be the number of all such distinct
pairs whose edge class is ${\cal R}_l$ (obviously
$\sum_{l=1}^M{t_l}=k'$). For example, when all the corresponding
pairs are distinct, we get $k'=k$ and $t_l = |\{ 1 \leq i \leq k :
{\cal R}(u_i,v_i) = {\cal R}_l \} |$. Notice that, by definition
of $S'$, for every $i$, $v_i \neq u_i + \frac{1}{2}\nu n$, and
thus ${\cal R}(u_i,v_i)$ is an ordinary edge class. It follows
that:
\begin{equation}\label{edge_prob_in_s'}
\Pr[u v \in E(G^k)] = \prod_{l=1}^M \frac{\nu - t_l}{\nu}
\end{equation}
This expression is minimal when $t_l=k'$ for some $l$, since
replacing $t_{l_1},t_{l_2}>0$ with
$t'_{l_1}=t_{l_1}+t_{l_2},t'_{l_2}=0$ strictly decreases its
value. Therefore $\Pr[u v \notin E(G^k)] \leq \frac{k'}{\nu} \leq
\frac{k}{\nu}$. Notice that, crucially, by the structure of $S'$,
as each edge class appears in at most one pair of vertices of
$S'$, the events $u v \notin E(G^k)$ are independent for different
pairs $u,v$. Let $A_{S'}$ denote the event that there is an
independent set $S'$ of the above form of size $s'=\lceil
c'\log_2(N) \rceil$, where $c'=2k^2$. Applying the same
consideration used on $S$ and $G$ to $S'$ and $G^k$, gives
(assuming $N$ is sufficiently large):
$$\Pr[A_{S'}] \leq \binom{N^k}{s'} \left(\frac{k}{\nu}\right)^
{\binom{s'}{2}} \leq N^{k
s'}2^{-\frac{1}{2}s'^2\log_2(\frac{\nu}{k})} \leq 2^{\left(k
c'-\frac{1}{2}\log_2(\frac{\nu}{k})c'^2\right)\log_2^2(N)}
$$
Now, our choice of $c'$ should satisfy $c' >
\frac{2k}{\log_2(\frac{\nu}{k})}$ for this probability to tend to
zero. Whenever $2 \leq k \leq \frac{\nu}{2}$ we get $k
\log_2(\frac{\nu}{k}) \geq k > 1$, thus $c'=2k^2 > \frac{2
k}{\log_2(\frac{\nu}{k})}$. For $\frac{\nu}{2} < k < \nu$ we have
$1 < \frac{\nu}{k} < 2$ and thus $\log_2(\frac{\nu}{k}) >
\frac{\nu}{k}-1$. Taking any $c' \geq \frac{2k^2}{\nu-k}$ would be
sufficient in this case, hence $c'=2k^2$ will do. Overall, we get
that $\Pr[A_{S'}]$ tends to 0
as $N$ tends to infinity. \\
\noindent Altogether, we have shown that for every $2 \leq k <
\nu$:
$$\alpha(G^k) \leq k^2 \nu \alpha(G^{k-1}) 2k^2 \log_2(N) = 2k^4
\nu \log_2(N) \alpha(G^{k-1})$$
Hence, plugging in the facts that
$\alpha(G)\leq 2\log_\nu(N) < 2\log_2(N)$ and $2^{\frac{m}{2}}m!
\leq m^m$ for $m\geq 2$, we obtain the following bound for all $k
\in \{1,\ldots,\nu-1\}$:
$$\alpha(G^k) \leq 2^k (k!)^4 \nu^{k-1} \log_2^k(N) \leq
2^{-k}k^{4k} \nu^{k-1} \log_2^k(N) $$
$$ a_k \leq \frac{1}{2}k^4 \nu \log_2(N) \leq \frac{1}{2}\nu^5 \log_2(N)$$
It remains to show that $a_\nu$ is large. Consider the following
set of vertices in $G^\nu$ (with addition modulo N):
\begin{equation}\label{indep-set} I = \{
~\overline{x}=(x,x+n,\ldots,x+(\nu-1)n)~|~0 \leq x < N~\}
\end{equation} Clearly $I$ is independent, since for any $0 \leq x
< y < N$, the corresponding coordinates of
$\overline{x},\overline{y}$ form one complete edge class, thus
exactly \textit{one} of these coordinates is disconnected in $G$.
This implies that $a_\nu \geq N^{\frac{1}{\nu}}$.

\noindent Hence, we have shown that for every value of $\nu$,
there exists a graph $G$ on $N$ vertices such that:
\begin{equation}\label{jmp-graph}
\left\{\begin{array}{l}
  a_i \leq c_i \log_2(N) ~~~(i=1,\ldots,\nu-1)\\
  a_\nu \geq N^{\frac{1}{\nu}} \\
\end{array}\right.
\end{equation}
\qed

\noindent We note that a simpler construction could have been
used, had we wanted slightly weaker results, which are still
asymptotically sufficient for proving the theorem. To see this,
take $N=n \nu$ and start with the complete graph $K_N$. Now order
the $N$ vertices arbitrarily in $n$ rows (each of length $\nu$),
as $(v_{i j})$ ($1 \leq i \leq n$, $1 \leq j \leq \nu$). For each
pair of rows $i,i'$, choose (independently) a single column $1\leq
j \leq \nu$, and remove the edge $v_{i j} v_{i' j}$ from the
graph. This gives a graph $G$ with $\binom{N}{2} - \binom{n}{2}$
edges. A calculation similar to the one above shows that with high
probability $a_k \leq c_k \log_2(N)$ for $k < \nu$, and yet
$\alpha(G^\nu) \geq n$ (as opposed to $N$ in the original
construction), hence $a_\nu \geq
\left(\frac{N}{\nu}\right)^{\frac{1}{\nu}} \geq \frac{1}{2}
N^{\frac{1}{\nu}}$.

\section{Graphs with an irregular independence series}
\label{sec::thm-two-proof} \noindent Theorem
\ref{thm-multiple-jumps} states that there exists a graph $G$ whose
independence series exhibits an arbitrary (finite) number of jumps.
Our first step towards proving this theorem is to examine the
behavior of fixed powers of the form $k \geq \nu$ for the graphs
described in the previous section. We show that these graphs, with
high probability, satisfy $a_k = \left(1+O(\log N)\right)
N^{\lfloor\frac{k}{\nu}\rfloor\frac{1}{k}}$, for every fixed $k \geq
\nu$. The notation $a_k=\left(1+O(\log N)\right)N^\alpha$, here and
in what follows, denotes that $N^\alpha \leq a_k \leq c N^\alpha
\log N$ for a fixed $c>0$. The lower bound of
$N^{\lfloor\frac{k}{\nu}\rfloor\frac{1}{k}}$ for $a_k$ can be
derived from the cartesian product of the set $I$, defined in
\eqref{indep-set}, with itself, $\lfloor\frac{k}{\nu}\rfloor$ times;
the upper bound is more interesting. Fix an arbitrary set $S$, as
before, however, this time, prior to generating $S'$, we first
remove from $S$ all vertices which contain among their coordinates a
set of the form $\{x,x+n,\ldots,x+(\nu-1)n\}$. This amounts to at
most $\binom{k}{\nu} \nu! n \alpha(G^{k-\nu})$ vertices. This step
ensures that $S$ will not contain vertices that share a relation,
such as the one appearing in the set $I$ defined in
\eqref{indep-set}. However, an edge class may still be completely
contained in the coordinates of $u,v \in S$, in an interlaced form,
for instance: $u=(x,y+n,x+2n,\ldots,x+(\nu-1)n,\ldots)$ and
$v=(y,x+n,y+2n,\ldots,y+(\nu-1)n,\ldots)$. This will be
automatically handled in generating $S'$, since all vectors $v$ with
$x+t n$ in any of their coordinates are removed from $S'$ after
processing the vector $u$. Equation \eqref{edge_prob_in_s'} remains
valid, with $t_i < \nu$ for all $i$, however now me must be more
careful in minimizing its right hand side. We note that for every $0
< t_i,t_j < \nu-1$, setting $t'_i=\nu-1, t'_j=t_i+t_j-t'_i$ reduces
the product of $\frac{(\nu-t_i)(\nu-t_j)}{\nu^2}$. Therefore, again
denoting by $k'$ the number of distinct pairs of corresponding
coordinates, we obtain the following bound on the probability of the
edge $u v$:
\begin{equation}\label{edge_prob_k_geq_nu}\Pr[u v \in E(G^k)] \geq
\left(\frac{1}{\nu}\right)^{\lfloor \frac{k'}{\nu-1} \rfloor}
\frac{\nu - \left(k' \bmod(\nu-1)\right)}{\nu} \geq
\left(\frac{1}{\nu}\right)^{\frac{k'}{\nu-1}} \geq
\left(\frac{1}{\nu}\right)^{\frac{k}{\nu-1}}
\end{equation}
Thus:
$$\Pr[u v \notin E(G^k)] \leq
\mathrm{e}^{-\left(\frac{1}{\nu}\right)^{\frac{k}{\nu-1}}}$$ Now,
the same consideration that showed $\alpha(G)\leq 2\log_\nu(N)$
implies that any set $S'$ generated from $S$ in this manner, which
is of size $s' \geq 2 k \nu^{ \frac{k}{\nu-1}} \log(N)$, is almost
surely not independent (for the sake of convenience, we set $p =
\mathrm{e}^{-\left(\frac{1}{\nu}\right)^{\frac{k}{\nu-1}}}$).
Indeed, the probability that there is such an independent set $S'$
is at most:
$$\binom{N^k}{s'}p^{\binom{s'}{2}} \leq
\frac{p^{-s'/2}}{s'!} \left(N^k p^{\frac{s'}{2}} \right)^{s'} =
\frac{p^{-s'/2}}{s'!}
\exp\left(k\log(N)-\frac{s'}{2}\nu^{-\frac{k}{\nu-1}} \right)^{s'}
\leq \frac{p^{-s'/2}}{s'!} = o(1)$$

\noindent Thus, almost surely, $|S'| \leq 2 k \nu^{ \frac{k}{\nu-1}}
\log(N)$. Altogether, we have:
$$\alpha(G^k) \leq \binom{k}{\nu} \nu! n \alpha(G^{k-\nu}) + k^2\nu
\alpha(G^{k-1}) \cdot 2k \nu^{\frac{k}{\nu-1}} \log(N)
 = $$
\begin{equation}\label{indp_bound_gt_nu} = \binom{k}{\nu} (\nu-1)! N \alpha(G^{k-\nu}) + 2k^3
\nu^{1+\frac{k}{\nu-1}}\log(N)\alpha(G^{k-1}) \end{equation}
 For $k=\nu$ and a sufficiently large $N$, we get
\begin{equation}\label{indp_bound_induction_base}
N \leq \alpha(G^\nu) \leq N (\nu-1)! +
2\nu^{5+\frac{1}{\nu-1}}\log(N)
\left(c_{\nu-1}\log_2(N)\right)^{\nu-1} \leq N\log_2(N)
\end{equation}
Set $d_1=\ldots=d_{\nu-1}=0$, $d_\nu=1$ and $d_k=4k^3
\nu^{1+\frac{k}{\nu-1}} d_{k-1}$ for $k>\nu$. It is easy to verify
that $\frac{1}{2}d_k \geq \binom{k}{\nu}(\nu-1)! d_{k-\nu}$, and
$\frac{1}{2}d_k \geq 2k^3 \nu^{1+\frac{k}{\nu-1}} d_{k-1}$. Hence,
by induction, equations \eqref{indp_bound_gt_nu} and
\eqref{indp_bound_induction_base} imply that for all $k \geq \nu$:
 $$
\alpha(G^k) \leq d_k N^{\lfloor\frac{k}{\nu}\rfloor} \log_2^k(N) $$
By definition of the $d_k$ series,
$$ d_k \leq 4^{k-\nu}\left(\frac{k!}{\nu!}\right)^3
\nu^{(k-\nu)\left(1+\frac{k}{\nu-1}\right)} \leq 4^k (k!)^3
\nu^{k\left(1+\frac{k-\nu}{\nu-1}\right)} = 4^k (k!)^3 \nu^{k
\frac{k-1}{\nu-1}}$$ Hence,
$$ 1 \leq
\frac{a_k}{N^{\lfloor\frac{k}{\nu}\rfloor \frac{1}{k}}} \leq
\sqrt{2}k^3 \nu^{\frac{k-1}{\nu-1}} \log_2(N) $$ as required.

\noindent Let us construct a graph whose independence series
exhibits two jumps (an easy generalization will provide any finite
number of jumps). Take a random graph, $G_1$, as described above,
for some index $\nu_1$ and a sufficiently large number of vertices
$N_1$, and another (independent) random graph, $G_2$ for some other
index $\nu_2 > \nu_1$, on $N_2 = N_1^{\alpha\frac{\nu_2}{\nu_1}}$
vertices (when $\alpha > 1)$. Let $G=G_1 \cdot G_2$ be the strong
product of the two graphs; note that $G$ has $N=N_1 N_2$ vertices.
It is crucial that we do not take $G_1$ and $G_2$ with jumps at
indices $\nu_1,\nu_2$ respectively \textit{separately}, but instead
consider the product $G$ of two random graphs constructed as above.
We claim that with high probability, $G$ satisfies:
$$a_k(G)=\left\{
\begin{array}{cl}
O(\log N) & k < \nu_1 \\
\left(1+O(\log N)\right) N_1^{\lfloor\frac{k}{\nu_1}\rfloor\frac{1}{k}}
& \nu_1 \leq k < \nu_2 \\
\left(1+O(\log N)\right)
N_2^{\lfloor\frac{k}{\nu_2}\rfloor\frac{1}{k}} & k \geq \nu_2 \\
\end{array} \right.$$
i.e., for $\nu_1 \leq k < \nu_2$ which is a multiple of $\nu_1$, we
have $a_k = \left(1+O(\log
N)\right)N^{\frac{1}{\nu_1+\alpha\nu_2}}$; for $k \geq \nu_2$ which
is a multiple of $\nu_2$ we have $a_k = \left(1+O(\log
N)\right)N^{\frac{\alpha}{\nu_1+\alpha\nu_2}}$. Therefore, we get an
exponential increase of order $\alpha$ at the index $\nu_2$, and
obtain two jumps, as required.

\noindent To prove the claim, argue as follows: following the
formerly described methods, we filter an arbitrary set $S \subset
V(G^k)$ to a subset $S'$, in which every two vertices have a
positive probability of being connected, and all such events are
independent. This filtering is done as before - only now, we
consider the criteria of both $G_1$ and $G_2$ when we discard
vertices. In other words, if we denote by $u^1,u^2$ the $k$-tuples
corresponding to $G_1^k$ and $G_2^k$ of a vertex $u\in V(G^k)$, then
a vertex $u \in S$ filters out the vertex $v$ from $S$ iff $u^1$
would filter out $v^1$ in $G_1^k$ or $u^2$ would filter out $v^2$ in
$G_2^k$ (or both). Recall that, by the method $S'$ is generated from
$S$, no two vertices in $S'$ share an identical $k$-tuple of $G_1^k$
or of $G_2^k$. Hence, two vertices $u,v \in S'$ are adjacent in
$G^k$ iff they are adjacent both in $G_1^k$ and in $G_2^k$. These are
two independent events, thus, by \eqref{edge_prob_in_s'} and
\eqref{edge_prob_k_geq_nu}, we get the following fixed lower bound
on the probability of $u$ and $v$ being adjacent:
$$\Pr[u v \in E(G^k)] = \Pr[u^1 v^1 \in E(G_1^k)]
\Pr[u^2 v^2 \in E(G_2^k)] \geq \Omega(1) $$ This provides a bound of
$O(\log N)$ for the size of $S'$. Combining this with the increase
in the values of $\{a_k\}$ at indices $\nu_1$ and $\nu_2$
($a_{\nu_i} \geq N_i^{\frac{1}{\nu_i}}$ for $i=1,2$) proves our
claim.

\noindent In order to obtain any finite number of jumps, at indices
$\nu_1,\ldots,\nu_s$, simply take a sufficiently large $N_1$ and set
$N_i = N_{i-1}^{\alpha \frac{\nu_i}{\nu_{i-1}}}$ for $1 < i \leq s$,
where $\alpha > 1$. By the same considerations used above, with high
probability the graph $G=G_1 \cdot \ldots \cdot G_s$ (where $G_i$ is
a random graph designed to have a jump at index $\nu_i$ almost
surely) satisfies $a_{\nu_i} \geq a_k^\alpha$ for all $k < \nu_i$.
Hence for every $\epsilon > 0$ we can choose
$\alpha>\frac{1}{\epsilon}$ and a sufficiently large $N_1$ so that
$a_k < a_{\nu_i}^\epsilon$ for all $k < \nu_i$. This completes the
proof.\qed

\section{Concluding remarks and open problems}\label{sec::final}
We have shown that even when the independence series stabilizes for
an arbitrary (fixed) number of elements, or jumps and then
stabilizes, it still does not necessarily approximate the Shannon
capacity up to any power of $\epsilon>0$.  However, our
constructions require the number of vertices to be exponentially
large in the values of the jump indices $\nu_i$. We believe that
this is not a coincidence, namely a prefix of \textit{linear} (in
the number of vertices) length of the independence series can
provide a good approximation of the Shannon capacity. The following
two specific conjectures seem plausible: {\conjecture For every
graph $G$ on $n$ vertices, $\max \{a_k\}_{k \leq n} \geq \frac{1}{2}
c(G)$, that is, the largest of the first $n$ elements of the
independence series gives a 2-approximation for $c(G)$.}
{\conjecture For every $\epsilon>0$ there exists an $r=r(\epsilon)$
such that for a sufficiently large $n$ and for every graph $G$ on
$n$ vertices, the following is true: $\max\{a_k\}_{k \leq n^r} \geq
(1-\epsilon)c(G)$. }

Our proof of Theorem \ref{thm-one-jump} shows the existence of a
graph whose independence series increases by a factor of $N^\delta$
at the $k$-th power, where $\delta=\frac{1-o(1)}{k}$. It would be
interesting to decide if there is a graph satisfying this property
for a \textit{constant} $\delta > 0$ (independent of $k$). This
relates to a question on channel discrepancy raised in
\cite{NogaRepeatedComm}, where the authors show that the ratio
between the independence number and the Shannon capacity of a graph
on $n$ vertices can be at least $n^{\frac{1}{2}-o(1)}$, and ask
whether this is the largest ratio possible. Proving Theorem
\ref{thm-one-jump} for a constant $\delta>0$ will give a negative
answer for the following question, which generalizes the channel
discrepancy question mentioned above: {\question Does $\max
\{a_i\}_{i \leq k}$, for any fixed $k\geq 2$, approximate $c(G)$ up
to a factor of $n^{\frac{1}{k}+o(1)}$ (where $n=|V(G)|$)? }

\noindent Although our results exhibit the difficulty in
approximating the Shannon capacity of a given graph $G$, this
problem is not even known to be NP-hard (although it seems plausible
that it is in fact much harder). We conclude with a question
concerning the complexity of determining the value of $c(G)$
accurately for a given graph $G$: {\question Is the problem of
deciding whether the Shannon Capacity of a given graph exceeds a
given value decidable? }

\end{document}